\documentclass[12pt]{article}
\usepackage{amssymb}
\usepackage{amsmath}
\usepackage{stmaryrd}
\usepackage{bbm}
\usepackage{graphicx}
\catcode`@=11 
\renewcommand{\section}{\@startsection{section}{1}{0pt}{\medskipamount}
{\medskipamount}{\bf}}
\numberwithin{equation}{section}
\catcode`@=12 
\def\a{\alpha}
\def\b{\beta}

\def\de{\delta}
\def\e{\epsilon}

\def\g{\gamma}
\def\p{\psi}

\def\r{\rho}
\def\s{\sigma}

\newcommand{\R}{\mathbb R}

\newcommand{\cN}{{\cal N}}
\newcommand{\unity}{\mathbbm{1}}
\def\be{\begin{equation}}
\def\ee{\end{equation}}
\def\arr{\begin{array}{rll}}
\def\ea{\end{array}}
\def\bea{\begin{eqnarray}}
\def\eea{\end{eqnarray}}
\def\sfrac#1#2{{\textstyle\frac{#1}{#2}}}

\def\ic{{\rm i}}

\def\pa{\partial}
\def\>{\rangle}
\def\<{\langle}
\def\+{\dagger}
\def\={\ =\ }
\def\und{\qquad\textrm{and}\qquad}
\def\ax{\a(x)}
\def\bx{\b(x)}
\def\gx{\g(x)}
\def\Mt{{M^{\smash{\!\!\top}}\!}}
\begin{document}
\renewcommand{\thefootnote}{\fnsymbol{footnote}}
\begin{titlepage}
\setcounter{page}{0}
\begin{flushright}
ITP--UH--13/08
\end{flushright}
\vskip 1cm
\begin{center}
{\LARGE\bf WDVV solutions from orthocentric \\[8pt]
           polytopes and Veselov systems \footnote{
\hbox{Contribution to ``Modern Problems in Theoretical Physics''
for the 60th birthday of Ioseph L.\ Buchbinder}} 
}
\vskip 2cm
$
\textrm{\Large Olaf Lechtenfeld}
$
\vskip 0.7cm
{\it
Institut f\"ur Theoretische Physik, Leibniz Universit\"at Hannover,\\
Appelstrasse 2, D-30167 Hannover, Germany} \\
{Email: lechtenf@itp.uni-hannover.de}
\vskip 0.2cm
\end{center}
\vskip 2cm
\begin{abstract} \noindent
$\cN{=}4$ superconformal $n$-particle quantum mechanics on the real line
is governed by two prepotentials, $U$ and $F$, which obey a system of
partial nonlinear differential equations generalizing the 
Witten-Dijkgraaf-Verlinde-Verlinde (WDVV) equation. 
For $U{=}0$ one remains with the WDVV equation which suggests an ansatz
for~$F$ in terms of a set of covectors to be found. One approach constructs
such covectors from suitable polytopes, another method solves Veselov's
$\vee$-conditions in terms of deformed Coxeter root systems. I relate the
two schemes for the $A_n$ example.
\end{abstract}
\vspace{2cm}
\end{titlepage}
\renewcommand{\thefootnote}{\arabic{footnote}}
\setcounter{footnote}0

\section{Introduction}
\noindent 
The issue of constructing $\cN{=}4$~superconformal extensions
of Calogero-type multi-particle quantum mechanics in one dimension
has been attacked in several works~\cite{wyl}--\cite{glp3}.
In \cite{wyl,bgl} it was discovered that this task leads to the
(generalized) Witten-Dijkgraaf-Verlinde-Verlinde (WDVV) equation known from
two-dimensional topological field theory~\cite{w,dvv}.
A physicist's classification of $\cN{=}4$~superconformal mechanics models 
based on particular WDVV~solutions has been advanced in~\cite{glp2,glp3},
where new models (with a purely quantum potential based on orthocentric
simplices) were found. Independently, mathematicians' efforts revealed
WDVV~solutions derived from Coxeter systems and certain deformations thereof
and lead to the notion of Veselov $\vee$-systems~\cite{magra}--\cite{feives2}.
In the current paper I relate the mathematics approach with the physicist's 
picture for solving the (generalized) WDVV~equation. In particular, 
the deformed $A_n$ solutions of~\cite{chaves} will be mapped to the 
orthocentric simplices of~\cite{glp3}.

In section~2 I recall the formulation $\cN{=}4$ superconformal 
$n$-particle mechanics in terms of $su(1,1|2)$ generators.
The closure of the superconformal algebra poses constraints on 
the interaction, which for an ansatz quartic in the fermionic coordinates
lead to the WDVV~equation plus a homogeneity condition for a quantum
prepotential~$F$ and to related differential equations for a classical
prepotential~$U$.
Section~3 expresses these prepotentials in terms of a system of covectors,
thereby turning the differential to nonlinear algebraic equations.
Putting $U$ to zero, a family of WDVV~solutions is constructed in section~4. 
Its covectors deform the $A_n$ root system and are parametrized by the 
shape moduli of orthocentric $n$-simplices.
The different formulations of the WDVV~equation are related in section~5,
where the geometry of the deformed $A_n$ $\vee$-systems is made explicit.
\vspace{1cm}

\section{WDVV equations from N=4 superconformal quantum mechanics}
\noindent
Let me consider a quantum mechanical system of $n$ identical particles 
with unit mass on the real line, described by positions~$x^i$ and
momenta $p_i$, and enhanced by fermionic degrees of freedom
$\psi^i_\a$ and $\bar\psi^{i\a}=(\psi^i_\a)^\+$, 
where $i=1,\dots,n$ and $\a=1,2$.
Spinor indices are raised and lowered with the invariant
tensor $\e^{\a\b}$ and its inverse $\e_{\a\b}$, where $\e^{12}=1$.
Further, I impose the canonical quantization rules\footnote{
I suppress $\hbar$ except for illustrative purposes.}
\be \label{quant}
[x^i, p_j]\=\ic  {\de_j}^i \und
\{\p^i_\a, {\bar\p}^{j\b} \}\=\,{\de_\a}^\b \de^{ij}\ ,
\ee
with all other (anti)commutators vanishing.
At this stage I have introduced a Euclidean metric $(\de_{ij})$
in the configuration space~$\R^{n|4n}/S_n$.

I want the dynamics to be invariant under $\cN{=}4$ superconformal 
transformations. Their generators
$\{H,Q_\a,\bar Q^\a,D,J_a,S_\a,\bar S^\a,K\}$, with $a=1,2,3$ 
and ${(Q_\a)}^\+{=}{\bar Q}^\a$ as well as ${(S_\a)}^\+{=}{\bar S}^\a$,
form a (centrally extended) $su(1,1|2)$ algebra defined by
the following non-vanishing (anti)commutation relations,
\begin{align}\label{algebra}
&
[D,H] \= -\ic \, H\ , && 
[H,K] \= 2\ic \, D\ ,
\nonumber\\[4pt]
&
[D,K] \= +\ic \, K\ , && 
[J_a,J_b] \= \ic \, \epsilon_{abc} J_c\ ,
\nonumber\\[2pt]
&
\{ Q_\a, \bar Q^\b \} \= 2\, H {\de_\a}^\b\ , &&
\{ Q_\a, \bar S^\b \} \=
+2\ic\,{{(\s_a)}_\a}^\b J_a-2\,D{\de_\a}^\b-\ic\,C{\de_\a}^\b\ ,
\nonumber\\[2pt]
&
\{ S_\a\,,\, \bar S^\b \} \= 2\, K {\de_\a}^\b\ , &&
\{ \bar Q^\a, S_\b \} \=
-2\ic\,{{(\s_a)}_\b}^\a J_a-2\,D{\de_\b}^\a+\ic\,C{\de_\b}^\a\ ,
\nonumber\\[2pt]
& 
[D,Q_\a] \= -\sfrac{1}{2} \ic\, Q_\a\ , && 
[D,S_\a] \= +\sfrac{1}{2} \ic \, S_\a\ ,
\nonumber\\[4pt]
&
[K,Q_\a] \= +\ic \, S_\a\ , && 
[H,S_\a] \= -\ic \, Q_\a\ ,
\nonumber\\[2pt]
&
[J_a,Q_\a] \= -\sfrac{1}{2} \, {{(\s_a)}_\a}^\b Q_\b\ , && 
[J_a,S_\a] \= -\sfrac{1}{2} \, {{(\s_a)}_\a}^\b S_\b\ ,
\nonumber\\[4pt]
& 
[D,\bar Q^\a] \= -\sfrac{1}{2} \ic \, \bar Q^\a\ , && 
[D,\bar S^\a] \= +\sfrac{1}{2} \ic \, \bar S^\a\ ,
\nonumber\\[4pt]
& 
[K,\bar Q^\a] \= +\ic \, \bar S^\a\ , && 
[H,\bar S^\a] \= -\ic \, \bar Q^\a\ ,
\nonumber\\[2pt]
&
[J_a,\bar Q^\a] \= \sfrac{1}{2} \, \bar Q^\b {{(\s_a)}_\b}^\a\ , && 
[J_a,\bar S^\a] \= \sfrac{1}{2} \, \bar S^\b {{(\s_a)}_\b}^\a\ .
\end{align}
Here, $\e_{123}=1$, $C$ stands for the central charge, 
and $\{\s_1,\s_2,\s_3\}$ denote the Pauli matrices.

For a realization of the generators I try
(repeated indices are summed over)~\cite{wyl}--\cite{glp3}
\be \label{realization}
\begin{aligned}
K &\= \sfrac{1}{2} x^i x^i \ , \qquad
S_\a \= x^i \p^i_\a\ , \qquad
\bar S^\a \= x^i \bar\p^{i\a}\ ,
\\[4pt]
D &\= -\sfrac{1}{4} (x^i p_i +p_i x^i) \ , \qquad
J_a \= \sfrac{1}{2} \bar\p^{i\a} {{(\s_a)}_\a}^\b \p^i_\b\ ,
\\[4pt]
Q_\a &\= \bigl(p_j-\ic\,x^i\,U_{ij}(x)\bigr)\,\p_\a^j \ \,-
\sfrac{\ic}{2}\,x^i\,F_{ijkl}(x)\,\<\p^j_\b\,\p^{k\b}\bar\p^l_\a\> \ ,\\[4pt]
\bar Q^\a &\= \bigl(p_j+\ic\,x^i\,U_{ij}(x)\bigr)\,\bar\p^{j\a} \,-
\sfrac{\ic}{2}\,x^i\,F_{ijkl}(x)\,\<\p^{j\a}\bar\p^{k\b}\bar\p^l_\b\>\ ,
\\[4pt]
H &\= \sfrac{1}{2} p_i p_i\ +\ V_B(x)\ -\
U_{ij}(x) \langle \p^i_\a {\bar\p}^{j\a} \rangle\ +\
\sfrac14 F_{ijkl}(x) \langle\p^i_\a\p^{j\a}\bar\p^{k\b}\bar\p^l_\b\rangle\ ,
\end{aligned}
\ee
with completely symmetric unknown functions $V_B$, $U_{ij}$ and $F_{ijkl}$
homogeneous of degree $-2$ in $x\equiv\{x^1,\ldots,x^n\}$.
Here, the symbol $\langle\dots\rangle$ stands for symmetric (or Weyl) ordering.
The ordering ambiguity present in the fermionic sector affects
the bosonic potential~$V_B$. In contrast to the $\cN{=}2$ superconformal
extensions~\cite{fm,glp1}, the closure of the algebra demands the quartic
term, and a nonzero central charge requires the quadratic term.
Hence, there does not exist a free mechanical representation of 
the algebra~(\ref{algebra}). A prototypical model is of the Calogero type,
\be 
V_B \= \sum_{i<j} \sfrac{g^2}{(x^i-x^j)^2} \ , \qquad
U_{ij} \= ? \ , \qquad F_{ijkl} \= ? \ .
\ee

Inserting the representation (\ref{realization}) into the 
algebra~(\ref{algebra}), one produces a fairly long list of constraints 
on $V_B$, $U_{ij}$ and $F_{ijkl}$. One of the consequences is 
that~\cite{wyl,bgl,glp2}
\be\label{pot}
\begin{aligned}
U_{ij} &\= \pa_i\pa_j U \und F_{ijkl} \= \pa_i\pa_j\pa_k\pa_l F\ ,\\[4pt]
V_B &\= \sfrac12\,(\pa_iU)(\pa_iU) \ +\
\sfrac{\hbar^2}8\,(\pa_i\pa_j\pa_kF)(\pa_i\pa_j\pa_kF) \ ,
\end{aligned}
\ee
which introduces two scalar prepotentials. 
Note that a quadratic polynomial in $F$ or a constant in $U$
are irrelevant.
The constraints then turn into the following system of 
nonlinear partial differential equations~\cite{bgl,glp2}
(see also~\cite{wyl}),
\bea
&&
(\pa_i\pa_k\pa_p F)(\pa_j\pa_l\pa_p F)\=
(\pa_j\pa_k\pa_p F)(\pa_i\pa_l\pa_p F)\quad,\qquad
x^i \partial_i \partial_j \partial_k F\=-\de_{jk}\ ,
\label{w1}
\\[6pt]
&&
\pa_i\pa_j U -(\pa_i\pa_j\pa_k F)\,\pa_k U\=0\quad,\qquad
\qquad\qquad\qquad\ \  x^i \pa_i U\=-C\ ,
\label{w2}
\eea
which I refer to as the ``structure equations''.
Notice that these equations are quadratic in~$F$ but only linear in~$U$.
The first of~(\ref{w1}) is a kind of zero-curvature condition for a
connection~$\pa^3F$. It coincides with the (generalized) WDVV equation
known from topological field theory~\cite{w,dvv}. The first of~(\ref{w2})
is a kind of covariant constancy for~$\pa U$ in the $\pa^3F$ background.
Since its integrability implies the WDVV~equation projected onto~$\pa U$,
I call it the ``flatness condition''. 

The right equations in (\ref{w1}) and~(\ref{w2}) represent homogeneity 
conditions for $U$ and~$F$. They are are inhomogeneous with constants 
$\de_{jk}$ and $C$ (the central charge) on the right-hand side
and display an explicit coordinate dependence.
Furthermore, the second equation in (\ref{w1}) can be integrated twice,
arriving at
\be\label{w4}
(x^i\pa_i - 2) F \= -\sfrac12\,x^ix^i \und x^i\pa_i U \= -C\ .
\ee
where I used the freedom in the definition of $F$ to put the integration
constants -- a linear function on the right-hand side -- to zero.

There are some dependencies among the equations (\ref{w1}) and~(\ref{w2}).
The contraction of two left equations with $x^i$ is a consequence of the
two right equations, and therefore only the components orthogonal to~$x$
are independent, effectively reducing the dimension to~$n{-}1$.
This means that only $\sfrac{1}{12}n(n{-}1)^2(n{-}2)$ WDVV equations 
need to be solved and only $\sfrac12n(n{-}1)$ flatness
conditions have to be checked. For $n{=}2$ in particular, the single
WDVV~equation follow from the homogeneity condition in~(\ref{w1}), and
the three flatness conditions are all equivalent. Hence, the nonlinearity
of the structure equations becomes only relevant for $n{\ge}3$.

\vspace{1cm}

\section{Covector ansatz for the prepotentials}
\noindent
For a particular solution to~(\ref{w4}), I make the ansatz~\cite{wyl,glp2,glp3}
\be\label{Fansatz}
F \= -\sfrac12\sum_{\a} f_\a\ \ax^2\,\ln|\ax| \und
U \= -\sum_{\a} g_\a\,\ln|\ax| 
\ee
with real coefficients $f_\a$ and~$g_\a$,
where $\a$ runs over a finite set of (unlabelled) noncollinear covectors 
in~$\R^n$, i.e.\
\be
\ax\=\a_i\,x^i \qquad\textrm{for each covector $\a$} \ .
\ee
The center-of-mass degree of freedom corresponds 
to $\ \ax=\r(x)\equiv\sum_ix^i$, and the relative particle motion is 
translation invariant only if $\ \a_i\r_i=0\quad\forall\a{\neq}\rho$, meaning 
that the other covectors span only the hyperplane perpendicular to~$\rho$
and $\{\a\}$ decomposes orthogonally. 
Identical particles require the set $\{\a\}$ to be invariant (up to sign) 
under permutations of the components~$\a_i$ and enforce equality of the
$f_\a$ (and $g_\a$) coefficients for permutation-related covectors.
Relative translation invariance and permutation symmetry are 
coordinate-dependent properties; they are not preserved
by a generic SO($n$) coordinate transformation. Therefore,
demanding either will severely restrict the coordinate choice.
Finally, a rescaling of~$\a$ may be absorbed into a renormalization of~$f_\a$.
Therefore, only the rays $\R_+\a$ are invariant data.
I cannot, however, change the sign of~$f_\a$ in this manner.

Compatibility of (\ref{Fansatz}) with the conditions~(\ref{w4}) directly yields
\be\label{hom}
\sum_\a f_\a\,\a_i \a_j \= \de_{ij} \und
\sum_\a g_\a \= C\ .
\ee
The second relation fixes the central charge, and the $g_\a$ are independent
free couplings if not forced to zero.
The first relation amounts to a decomposition of the identity $(\de_{ij})$
into (usually non-orthogonal) rank-one projectors and imposes 
$\sfrac12n(n{+}1)$ relations on the coefficients~$f_\a$ for a given 
set~$\{\a\}$.

{}From (\ref{Fansatz}) one derives
\be\label{YWform}
\pa_i\pa_j\pa_kF \= -\sum_{\a} f_\a\,\frac{\a_i \a_j \a_k}{\ax} \und
\pa_iU \= -\sum_{\a} g_\a\,\frac{\a_i}{\ax}\ ,
\ee
and so the bosonic part of the potential takes the form
\be
V_B \= \sfrac12 \sum_{\a,\b} \frac{\a{\cdot}\b}{\ax\;\bx}\,
\Bigl( g_\a g_\b\ +\ \sfrac{\hbar^2}4 f_\a f_\b\,(\a{\cdot}\b)^2 \Bigr)
\ee
with the covector scalar product
\be
\a{\cdot}\b \= \a_i\,\de^{ij}\b_j \= \a_i\,\b_i\ .
\ee
The remaining structure equations in (\ref{w1}) and~(\ref{w2}) become
\bea \label{FF} &&
\sum_{\a,\b}f_\a f_\b\,\frac{\a{\cdot}\b}{\ax\,\bx}\,(\a\wedge\b)^{\otimes2}\=0
\qquad\und \\[6pt] \label{UF} &&
\sum_{\b} \Bigl( g_\b\,\frac{1}{\bx}\ -\ f_\b \sum_\a g_\a\,
\frac{\a{\cdot}\b}{\ax} \Bigr)\,\frac{1}{\bx}\ \b\otimes\b \=0
\eea
with 
\be
(\a\wedge\b)^{\otimes2}_{ijkl} \= (\a_i\b_j-\a_j\b_i)(\a_k\b_l-\a_l\b_k)
\und (\b\otimes\b)_{ij} \= \b_i\,\b_j \ .
\ee

The task is to first solve (\ref{FF}) and~(\ref{hom}), 
i.e.~find sets~$\{\a,f_\a\}$,
and then to determine $\{g_\a\}$ from~(\ref{UF}), subject to~(\ref{hom}). 
Many $F$~backgrounds do not admit a $C{\neq}0$~solution,
but a homogeneous~$U$ can always be found~\cite{glp3}.
I close the section with a simplifying observation.
If a set of covectors decomposes into mutually orthogonal subsets, 
(\ref{FF}) and~(\ref{UF}) hold for each subset individually, 
and their prepotentials just add up to the total $F$ or~$U$.
Therefore, one may restrict the analysis to indecomposable covector sets. 
\vspace{1cm}

\section{WDVV solutions from orthocentric simplices}
\noindent
For the rest of the paper I put $U$ to zero and investigate solutions
to the WDVV equations~(\ref{FF}), subject to the homogeneity condition
\be \label{hcond}
\sum_\a f_\a\ \a\otimes\a \= \unity\ .
\ee
Let me look for indecomposable sets of covectors obeying the
WDVV equation~(\ref{FF}).
In one dimension, the equation is trivial.
For $n{=}2$, it follows from the homogeneity condition~(\ref{hcond}),
which can actually be satisfied
for {\it any\/} set~$\{\a\}$ of coplanar covectors~\cite{glp3}.
Nevertheless, it is instructive to outline the simplest examples.
For the case of two covectors $\{\a,\b\}$ one is forced to $\a{\cdot}\b=0$.
For three coplanar covectors $\{\a,\b,\g\}$,
the homogeneity condition~(\ref{hcond}) uniquely fixes the $f$~coefficients to
\be \label{f2dim}
f_\a \= -\frac{\b\cdot\g}{\a{\wedge}\b\ \g{\wedge}\a}
\qquad\textrm{and cyclic}\ ,
\ee
due to the identity
\be
\b{\wedge}\g\ \b{\cdot}\g\ \a^i\a^j\ +\ \textrm{cyclic} \= 
-\a{\wedge}\b\ \b{\wedge}\g\ \g{\wedge}\a\ \de^{ij}\ .
\ee
The traceless part of the homogeneity condition should imply
the single WDVV~equation~(\ref{FF}) in two dimensions. 
Indeed, the choice~(\ref{f2dim}) turns the latter into
\be
\a{\wedge}\b\ \gx\ +\
\b{\wedge}\g\ \ax\ +\
\g{\wedge}\a\ \bx\= 0 
\ee
which is identically true.
Without loss of generality I may assume that $\a+\b+\g=0$, i.e.~the three
covectors form a triangle. In this case I have
$\a{\wedge}\b=\b{\wedge}\g=\g{\wedge}\a=2A$, where the area~$A$ of the 
triangle may still be scaled to~$\sfrac12$, and (\ref{f2dim}) simplifies to
\be \label{f2sim}
f_\a \= -\frac{\b\cdot\g}{4\,A^2}
\qquad\textrm{and cyclic}\ .
\ee
\begin{figure}[ht]
\centerline{\includegraphics[width=8cm]{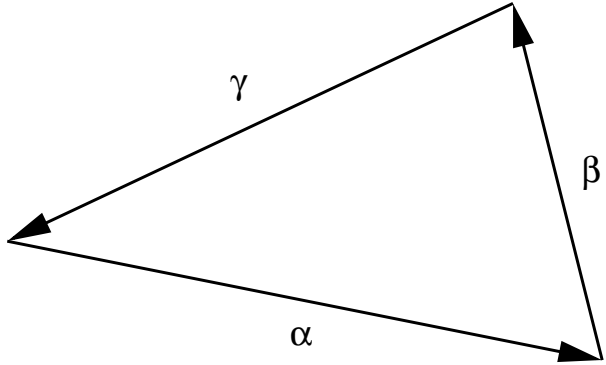}}
\caption{Triangular configuration of covectors}
\label{fig:1}
\end{figure}

In dimension $n{=}3$, the minimal set of three covectors must form
an orthogonal basis, with $f_\a^{-1}=\a{\cdot}\a$. Let me skip the cases
of four and five covectors and go to the situation of six covectors
because the homogeneity condition~(\ref{hcond}) then precisely determines
all $f$~coefficients.
However, it is not true that six generic covectors can be scaled to form 
the edges of a polytope. The space of six rays in~$\R^3$ modulo rigid~SO(3) 
is nine dimensional, while the space of tetrahedral shapes (modulo size) has
only five dimensions. In order to generalize the $n{=}2$ solution above,
let me assume that my six covectors can be scaled to form a tetrahedron,
with volume~$V$ and edges $\{\a,\b,\g,\a',\b',\g'\}$ where $\a'$ is skew 
to~$\a$ and so on. 
\begin{figure}[ht]
\centerline{\includegraphics[width=8cm]{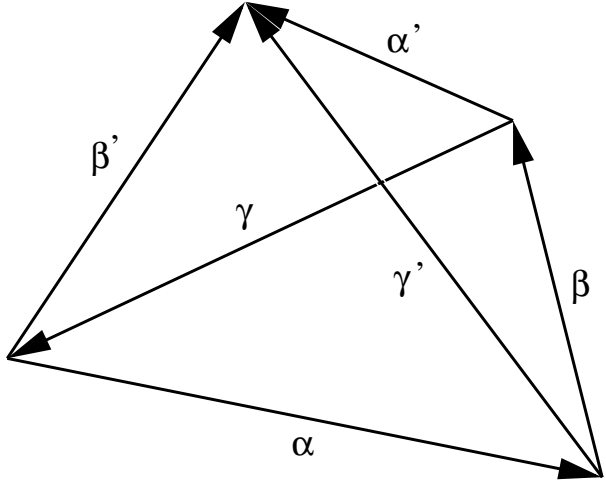}}
\caption{Tetrahedral configuration of covectors}
\label{fig:2}
\end{figure}
Any such tetrahedron is determined by giving three nonplanar
covectors, say $\{\a,\b,\g'\}$, which up to rigid rotation are fixed by
six parameters, corresponding to the shape and size of the tetrahedron.

The triangle result~(\ref{f2sim}) can be employed to patch together the unique
solution to the homogeneity condition~(\ref{hcond}) for the tetrahedron,
but only if the geometric constraints
\be \label{oc}
\a\cdot\a' \=0\ ,\qquad \b\cdot\b' \=0\ ,\qquad \g\cdot\g' \=0
\ee
are obeyed for the pairs of skew edges. In this situation, the identity
\be
\b{\cdot}\g\ \b'{\cdot}\g'\ \a^i\a^j\ +\
\b{\cdot}\g'\ \b'{\cdot}\g\ {\a'}^i{\a'}^j\ +\
\textrm{cyclic} \= -36\,V^2\,\de^{ij}\ ,
\ee
guarantees the homogeneity condition~(\ref{hcond}) for
\be \label{f3sim}
f_\a \=
-\frac{\b{\cdot}\g\ \b'{\cdot}\g'}{36\,V^2} \und
f_{\a'} \=
-\frac{\b{\cdot}\g'\ \b'{\cdot}\g}{36\,V^2}
\ee
plus their cyclic images.
Tetrahedra subject to~(\ref{oc}) are called ``orthocentric''~\cite{ehm}. 
They are characterized by
the fact that all four altitudes are concurrent (in the orthocenter) and
their feet are the orthocenters of the faces. The space of orthocentric
tetrahedra is of codimension two inside the space of all tetrahedra and 
represents a three-parameter deformation of the $A_3$~root system 
(ignoring the overall scale).

What about the WDVV~equation in this case?
The 15 pairs of edges in the double sum of~(\ref{FF}) group
into four triples corresponding to the tetrahedron's faces plus
the three skew pairs. It is not hard to see that for each face 
the contributions add to zero, and so the concurrent edge pairs
do not contribute to the double sum in~(\ref{FF}).
This leaves the three skew pairs, but their contribution is killed
by the orthocentricity constraint~(\ref{oc}), and the WDVV~equation
is indeed obeyed.

Although I do not know the $f$~coefficients for a general tetrahedron, 
I can offer the following proof that the WDVV~equation already enforces 
the orthocentricity. Consider the limit $\hat{n}(x)\to\infty$ for some 
fixed covector~$\hat{n}$ of unit length. Decomposing
\be \label{limit}
\a \= \a{\cdot}\hat{n}\;\hat{n} + \a_\perp \qquad\longrightarrow\qquad
\ax \= \a{\cdot}\hat{n}\;\hat{n}(x) + \a_\perp(x)
\ee
we see that any factor $\frac1\ax$ vanishes in this limit
unless $\a{\cdot}\hat{n}=0$. Thus, only covectors perpendicular to~$\hat{n}$ 
survive in (\ref{FF}) and~(\ref{UF}), reducing the system to the hyperplane
orthogonal to~$\hat{n}$. On the other hand, any solution to these equations,
being an identity in~$x$, must carry over to a solution of the limiting
equations, which correspond to the dimensionally reduced system.
In a general tetrahedron, take $\hat{n}\propto\a{\wedge}\a'$.
Then, the limit $\hat{n}(x)\to\infty$ in (\ref{FF}) retains only 
the covectors $\a$ and~$\a'$, and the WDVV equation reduces to a single term,
which vanishes only for $\a{\cdot}\a'=0$. Equivalently, the plane spanned by
$\a$ and $\a'$ contains no further covector, and two covectors in two
dimensions must be orthogonal. The same argument applies to
$\b{\cdot}\b'$ and $\g{\cdot}\g'$, completing the proof.

\begin{figure}[ht]
\centerline{\includegraphics[width=8cm]{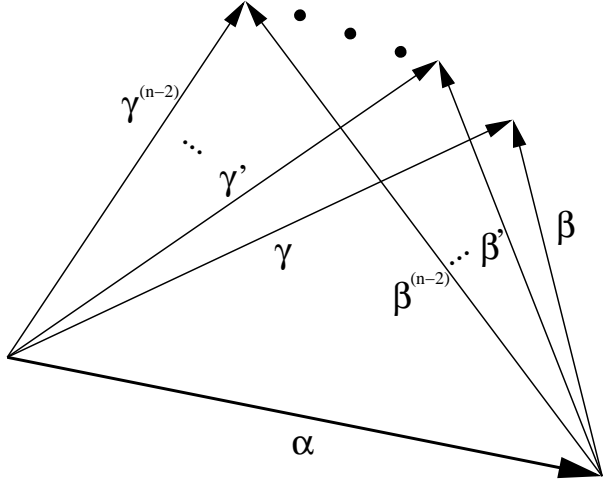}}
\caption{Faces sharing an edge of an $n$-simplex}
\label{fig:3}
\end{figure}
This scheme may be taken to any dimension~$n$. A simplicial configuration
of $\sfrac12n(n{+}1)$ covectors is already determined by $n$ independent
covectors, which modulo~SO($n$) are given by $\sfrac12n(n{+}1)$ parameters.
The homogeneity condition~(\ref{hcond}) uniquely fixes the $f$~coefficients.
Employing an iterated dimensional reduction to any plane spanned by a
skew pair of edges and realizing that no other edge lies in such a plane, one 
sees that the WDVV equation always demands such an edge pair to be orthogonal.
This condition renders the $n$-simplex orthocentric and reduces the 
number of degrees of freedom to~$n{+}1$ (now including the overall scale
given by the $n$-volume~$V$). In this situation I can write down 
the unique solution to both the homogeneity condition and the WDVV~equation,
\be
f_\a \= \frac{\b{\cdot}\g\ \b'{\cdot}\g'\ \b''{\cdot}\g''\
\cdots\ \b^{(n-2)}{\cdot}\g^{(n-2)}}{(n!\ V)^2}\ ,
\ee
where the edge~$\a$ is shared by the $n{-}1$ faces $\<\a\b\g\>$, 
$\<\a\b'\g'\>$, $\ldots$, $\<\a\b^{(n-2)}\g^{(n-2)}\>$, and I have
oriented all edges as pointing away from~$\a$.
This formula works because any sub-simplex, in particular any tetrahedral
building block, is itself orthocentric.
To summarize, the WDVV solutions for simplicial covector configurations
in any dimension are exhausted by an $n$-parameter deformation of the
$A_n$~root system. The $n$~moduli are relative angles and do not include the
$\sfrac12n(n{+}1)$ trivial covector rescalings, which, apart from the common
scale, destroy the tetrahedron. 

These findings suggest that covector configurations corresponding to 
deformations of other roots systems may solve the WDVV equations as well.
For verification, I propose to consider the polytopes associated with the 
weight systems of a given Lie algebra, since their edge sets are built from 
the root covectors. The idea is then to relax the angles of such polytopes 
and analyze the constraints from the homogeneity and WDVV~equations. 
The above $n$-dimensional orthocentric hypertetrahedra emerge simply from
the fundamental representations of $A_n$.
Extending this strategy to other representations and Lie algebras 
could lead to many more solutions.
\vspace{1cm}

\section{WDVV solutions from Veselov systems}
\noindent
In the mathematical literature, the (generalized) WDVV~equation is
usually formulated as
\be \label{wdvv1}
W_i\,W_k^{-1}W_j\=W_j\,W_k^{-1}W_i\qquad\textrm{for}\quad i,j,k=1,\ldots,n\ ,
\ee
where $W_i$ is an $n{\times}n$ matrix with entries
\be
(W_i)_{lm} \= \pa_i\pa_l\pa_m W \qquad\textrm{for}\quad W=W(y^1,\ldots,y^n)\ ,
\ee
and $\pa_i\equiv\sfrac{\pa}{\pa y^i}$. 
It is easy to show~\cite{magra} that (\ref{wdvv1}) is equivalent to
\be \label{wdvv2}
W_i\,G^{-1}W_j\=W_j\,G^{-1}W_i \qquad\textrm{with}\quad G=-y^k\,W_k\ ,
\ee
which in components reads
\be \label{wdvv3}
(\pa_i\pa_l\pa_p W)\,G^{pq}(\pa_q\pa_m\pa_j W)\= 
(\pa_j\pa_l\pa_p W)\,G^{pq}(\pa_q\pa_m\pa_i W) \ ,
\ee
where the index position distinguishes between the metric~$G$ and 
its inverse~$G^{-1}$. For the covector ansatz~(\ref{Fansatz})
\be \label{Wansatz}
W \= -\sfrac12\sum_{\b} f_\b\ \b(y)^2\,\ln|\b(y)|
\ee
it follows that
\be
W_i \= -\sum_{\b} f_\b\ \frac{\b_i}{\b(y)}\ \b\otimes\b 
\qquad\longrightarrow\qquad 
G \= \sum_{\b} f_\b\ \b\otimes\b \ .
\ee

How is this related to the material of the previous sections?
Comparing with (\ref{hcond}), it seems that one must impose
the additional condition of $G=-\unity$. However, this is not so, 
because such a choice may be achieved by a linear coordinate change
\be
x^i \= y^j\,M_j^{\ i} \qquad\longrightarrow\qquad \b_i \= M_i^{\ j}\,\a_j
\ee
so that for $\ F(x)=W(y)\ $ one gets
\be
W_i \= M_i^{\ j}\,F_j \und G_{lm} \= -y^k\,W_{klm} \= 
-M_l^{\ i}M_m^{\ j}\,x^k F_{kij} \= M_l^{\ i}\,\de_{ij}\,\Mt^j_{\ m}\ ,
\ee
where the right equation in~(\ref{w1}) was used in the last step.
This converts the metric~$(G_{ij})$ of the $y$-frame to the Euclidean 
metric~$(\de_{ij})$ in the $x$-frame,\footnote{
Note that for the $y$-frame one must replace $\de^{ij}$ with $G^{ij}$ in
the quantization rule~(\ref{quant}).}
and changes the covector scalar product accordingly,
\be
\b{\cdot}\b' \= \b_i\,G^{ij} \b'_j \= 
\a_k\,\Mt^k_{\ i}\,G^{ij} M_j^{\ l}\,\a'_l \= 
\a_k\,\de^{kl} \a'_l \= \a{\cdot}\a'\ ,
\ee
in short:
\be \label{factor}
G \= M\,\de\,\Mt \und \de \= \Mt\,G^{-1}\,M\ .
\ee
Thus, solutions to (\ref{wdvv3}) of the form~(\ref{Wansatz}) can be translated
to solutions to~(\ref{w1}) of the form~(\ref{Fansatz}) by a linear 
transformation.

For a prominent example, I turn to the $n$-parameter deformation of
the $A_n$ root system first proposed in~\cite{chaves},
\be \label{Andef}
\bigl\{ \b \bigr\} \= \bigl\{
\sqrt{c_ic_j}\,(e^i{-}e^j)\ ,\ \sqrt{c_i}\,e^i\ \bigm|\ 1\le i<j\le n\bigr\}
\qquad\textrm{(no sums)}\ ,
\ee
where $e^i(y)=y^i$ and the $c_i$ are arbitrary (positive) parameters. 
It was shown that this covector set satisfies the
so-called $\vee$-conditions, which implies that (with $f_\b=1$) it provides
an $n$-parameter family of solutions~(\ref{Wansatz}) to the WDVV~equation.
For this case, the metric and its inverse are quickly evaluated,
\be
G_{ij} \= \bigl( 1+{\textstyle\sum_k}c_k \bigr)\,c_i\,\de_{ij} - c_ic_j \und
G^{ij} \= \bigl( 1+{\textstyle\sum_k}c_k \bigr)^{-1} 
          \bigl( c_i^{-1}\de^{ij} + 1 \bigr)\ ,
\ee
but in order to compute the corresponding transformation matrix~$M$ 
(or its inverse~$M^{-1}$)
via~(\ref{factor}) one has to diagonalize $G$ (or $G^{-1}$), 
which is not an easy task.

However, in order to interpret the solution~(\ref{Andef}) in the $x$-frame,
it suffices to study its geometric (frame-independent) properties. First, 
I rescale each~$\b$ by shifting the square roots into $f_\b$~coefficients,
\be \label{simplex}
\bigl\{ \g \bigr\} \= \bigl\{ e^i{-}e^j\ ,\ e^i \bigr\} \und
\bigl\{ f_{\g} \bigr\} \= \bigl\{ c_ic_j\ ,\ c_i \bigr\} \qquad
\textrm{for}\quad 1\le i<j\le n\ ,
\ee
and observe that the new covectors fulfil the incidence relations of 
an $n$-simplex. Second, I must figure out the angles formed by its edges,
\be
\cos\angle(\g,\g') \= \frac{\g{\cdot}\g'}{\sqrt{\g{\cdot}\g\ \g'{\cdot}\g'}}
\qquad\textrm{with}\quad \g{\cdot}\g' \= \g_i\,G^{ij}\,\g'_j\ .
\ee
These angles depend on the deformation parameters~$c_i$, except for
\be
e^i\cdot(e^j{-}e^k) \= 0\qquad\textrm{for $i$, $j$, $k$ mutually distinct}\ ,
\ee
which means that non-concurrent edges are orthogonal to one another!
This is a frame-independent statement and qualifies the polytope based
on~(\ref{Andef}) as an orthocentric one. 

Clearly, I have rediscovered the solution family of section~4. 
As a side result, one obtains an explicit
parametrization of orthocentric $n$-simplices,
\be
\bigl\{ \a \bigr\}(c) \= \bigl\{ M^{-1}(e^i{-}e^j)\ ,\ M^{-1}e^i \bigr\} \ ,
\ee
where the $c_i$-dependence enters via the matrix~$M^{-1}$.
The (physical) geometries corresponding to the other known $\vee$-systems 
remain to be worked out.

\vspace{0.5cm}

\section*{Acknowledgments}
\noindent
I thank A.~Galajinsky and K.~Polovnikov for pleasant collaborations
on which this contribution is based. Furthermore, I am grateful to 
Misha Feigin for enlightening discussions.
The research was supported by DFG grant 436 RUS 113/669/0-3.

\end{document}